\documentclass[10pt,english,pre,nofootinbib,reprint,showpacs,showkeys,preprintnumbers,floatfix]{revtex4-1}
\usepackage[latin9]{inputenc}
\usepackage[a4paper]{geometry}
\usepackage{xcolor}
\usepackage{pdfcolmk}
\usepackage{babel}
\usepackage{units}
\usepackage{amsmath}
\usepackage{amssymb}
\usepackage{esint}
\PassOptionsToPackage{normalem}{ulem}
\usepackage{ulem}
\usepackage[unicode=true,pdfusetitle,
 bookmarks=true,bookmarksnumbered=false,bookmarksopen=false,
 breaklinks=false,pdfborder={0 0 1},backref=false,colorlinks=false]
 {hyperref}

\makeatletter

\providecolor{lyxadded}{rgb}{0,0,1}
\providecolor{lyxdeleted}{rgb}{1,0,0}
\DeclareRobustCommand{\lyxsout}[1]{\ifx\\#1\else\sout{#1}\fi}

\usepackage{aas_macros}

\usepackage{bbold}

\makeatother

\begin{document}

\title{Determining the composition of radio plasma via circular polarization:
\\
the prospects of the Cygnus A hot spots }

\begin{abstract}
The composition of the relativistic plasma produced in active galactic
nuclei and ejected via powerful jets into the interstellar/intergalactic
medium is still a major unsettled issue. It might be a positron-electron
plasma in case the plasma was created by pair production in the intense
photon fields near accreting super-massive black holes. Alternatively,
it might be an electron-proton plasma in case magnetic fields lift
and accelerate the thermal gas of accretion discs into relativistic
jets as the recent detection of $\gamma$-rays from blazars indicates.
Despite various attempts to unambiguously establish the composition
of the relativistic jets, this remains a major unknown. Here, we propose
a way to settle the question via sensitive measurements of circular
polarization (CP) in the radio emission of the hot spots of bright
radio galaxies like Cygnus A. The CP of synchrotron emission is determined
by the circular motions of the radiating relativistic leptons. In
case of charge symmetric energy spectra of an electron-positron plasma,
it should be exactly zero. In case of an electron-proton plasma the
electrons imprint their gyration onto the CP and we expect the hot
spots of Cygnus A to exhibit a fractional CP at a level of $10^{-3}(\nu/\mbox{GHz})^{-\nicefrac{1}{2}}$,
which is challenging to measure, but not completely unfeasible.
\end{abstract}

\author{Torsten A. Enßlin}

\affiliation{
Max Planck Institute for Astrophysics, Karl-Schwarzschildstr.1, 85741
Garching, Germany;\\
Ludwig-Maximilians-Universität München, Geschwister-Scholl-Platz 1,
80539 Munich, Germany}

\author{Sebastian Hutschenreuter}

\affiliation{
Max Planck Institute for Astrophysics, Karl-Schwarzschildstr.1, 85741
Garching, Germany;\\
Ludwig-Maximilians-Universität München, Geschwister-Scholl-Platz 1,
80539 Munich, Germany}

\author{Gopal-Krishna}

\affiliation{
Aryabhatta Research Institute of Observational Sciences (ARIES), Manora
Peak, Nainital - 263129, INDIA}

\maketitle

\section{Introduction}

\subsection{Circular polarization }

The radio synchrotron emission of relativistic plasma might exhibit
a slight circular polarization (CP) due to the circular motions of
the emitting leptonic particles, either electrons, or even electrons
and positrons (see below). The radiation of relativistic electrons
and positrons is beamed in the direction of their instantaneous motion
and for this reason very little of the circular motion of an ensemble
of relativistic leptons gets imprinted onto their radiation as circular
polarization. Nevertheless, an asymmetry in the charge distribution
of particles contributing to a given observational frequency should
leave a weak CP signal. Such an asymmetry should exist if the radio
plasma is mainly an electron-proton plasma, as one might expect in
case the radio jets consist of material launched magnetically from
accretion discs around super-massive black holes. On the other hand,
no CP signal should arise in case radio plasma is composed of electrons
and positrons created in similar amounts and with identical spectra
by pair production events from hard photons in the vicinity of such
black holes. Thereafter, acceleration and energy loss mechanism should
not distinguish between electrons and positrons in case the radio
plasma stays purely leptonic in its subsequent evolution. Then the
dynamics of electrons and positrons is (mirror) symmetric and they
should aquire similar energy spectra, which leads to cancellation
of their individual CP contributions.

Therefore, the detection of CP intrinsically emitted from radio plasma
would indicate that the charge symmetry of the synchrotron radiation
arising from the plasma constituents is broken, and therefore either
an electron proton plasma is dominant, or alternatively, the particle
acceleration mechanism favours electrons or positrons for a yet unknown
reason. 

More interesting, therefore, would be a reliable observational upper
bound on the CP flux from radio plasma, which is distinctly below
the expected level for an electron-proton plasma. That would be a
strong evidence against emission from an electron-proton plasma and
therefore favor an electron-positron plasma.

Other sources of CP should be excluded before accepting the electron-proton
plasma scenario via CP detection. The primary candidates for this
are Faraday conversion and instrumental polarization. 

Faraday conversion is the transformation of linearly polarized emission
into CP by the different dispersion a linear polarized wave experiences
depending on whether its polarization is parallel or perpendicular
to the magnetic field. It is believed to be the main cause of the
highly variable CP observed in compact objects \citep{1977ApJ...214..522J,1977ApJ...215..236J,1982ApJ...263..595H,1984Ap&SS.100..227V,1984MNRAS.208..409K,1990MNRAS.242..158B,2002A&A...388.1106B,2002ApJ...573..485R,2003A&A...401..499E}
like Sagrittarius\ $\text{A}^{*}$, GRS\ 1915, SS\ 433 \citep{1999ApJ...523L..29B,1999ApJ...526L..85S,2000ApJ...530L..29F,2002MNRAS.336...39F}.
Faraday conversion can be identified via its characteristic dependence
on wavelength to the third power, as long one does not probe the extreme
regimes like in AGNs \citep{2008IJMPD..17.1531V}. It operates most
efficiently in the strong magnetic fields and dense plasma present
near these objects. Thus, if we wish to minimise the contamination
to CP on account of Faraday conversion, we should investigate regions
with weaker magnetic field strengths and lower particle densities.
As we still need a high brightness, the hot spots of radio galaxies
seem to be ideal, and the brightest of those are the ones seen at
the tips of the radio lobes of Cygnus A, which will therefore be the
focus of the present theoretical study. 

Instrumental polarization is a severe problem, given that the fractional
CP is expected to be very small. Any uncorrected leakage of total
intensity or linear polarization into CP channels could lead to a
spurious detection. There are two possibilities to place limits on
the level of instrumental polarization. The first is that the observation
of many sources taken together should not show any preference for
any one sign of the CP, as this is determined by whether the magnetic
field in the emitting region is pointing towards us or away from us
and therefore should happen with equal frequency. Furthermore, the
temporal evolution of magnetic field structures in hot spots of radio
galaxies should be well beyond human timescales, so that a given hot
spot should always exhibit the same CP. Thus, observations with different
instruments having different systematics can allow us to gain confidence
about the genuineness of an inferred CP signal and its sign. Secondly,
in case Faraday rotation information for the hot spot plasma exists,
the expected CP sign in case of an electron proton plasma will be
known in advance. CP should exhibit a rotational sense that is opposite
to that of the direction of the Faraday rotation \citep{2017PhRvD..96d3021E}.
An agreement between the expected and observed CP signs would provide
confidence in the genuineness of the (intrinsic) CP signal, whereas
a disagreement would be indicative of uncorrected instrumental polarization.
Note, however, that an observed Faraday rotation most likely traces
magnetic fields surrounding the hot spot. If the Faraday rotation
is indeed intrinsic, then a charge asymmetry in at least the thermal
plasma there is confirmed, since such an asymmetry is a necessary
condition of the Faraday effect. 

\subsection{Previous studies}

Previous proposals to settle the question of the composition of radio
plasma were usually based on energetic arguments. More importantly,
from an observational standpoint, the composition of the relativistic
nonthermal plasma radiating in the jets and lobes of radio galaxies
continues to be a long-standing issue (e.g.,\citep{1998Natur.395..457W,2000ApJ...534..109S,1978PhRvL..41..135N,2006ApJ...648..200D}).
This question is crucial also because the acceleration and deceleration
of the jet near a supermassive black hole depends on the composition
of its plasma (e.g.,\citep{1977MNRAS.179..433B,1982MNRAS.199..883B,1982MNRAS.198.1109P}).
Over the years, a number of authors have argued in favour of the dominance
of relativistic electron-positron pair-plasma in the jets ejected
by the central engines of radio galaxies, both from observational
and theoretical perspectives (e.g.,\citep{1980Natur.288..149K,1996MNRAS.283..873R,1998Natur.395..457W,2000ApJ...545..100H}).
More recent arguments for a significant, if not dominant, presence
of relativistic pair plasma in the radio lobes include those put forward
by Kawakatu et al. \citep{2016MNRAS.457.1124K} and Potter \citep{2018MNRAS.473.4107P}
who also provide a summary of the observational and theoretical work
done on this topic. 

According to a currently popular scenario, the generation of relativistic
pair-plasma begins with photon-photon pair production in the spark
gap of the magnetosphere surrounding a supermassive black-hole (\citep{1977MNRAS.179..433B,1969ApJ...157..869G})
and the process continues along the length of the jet, e.g. via dissipation
of magnetic energy by reconnection (see, e.g., \citep{2018MNRAS.473.4107P}).
Although, at large distances from the black-hole, some hadronic plasma
may gradually get entrained in the jet flow, this process is unlikely
to be efficient at least in Fanaroff-Riley type II (FR II) radio galaxies,
since their jets remain effectively shielded by the cocoon of relativistic
lobe plasma \citep{2018MNRAS.473.4107P,2004ApJ...606..804M,1974MNRAS.166..513S}.
For FR I radio galaxies, the need for non-radiating pressure support
of the radio plasma hints at an efficient entrainment of protons into
the jet \citep{2018MNRAS.476.1614C}. 

A hadronic jet model was proposed by Mannheim and collaborators \citep{1996SSRv...75..331M,1998Sci...279..684M,1991A&A...251..723M,1992A&A...253L..21M}
to explain the TeV gamma-ray emission of blazars. The same emission
can, however, also be explained in terms of leptonic models, which
invoke synchrotron self-Compton (SSC) emission to produce TeV photons. 

In the case of FR II radio sources, hadronic jets have been considered
to explain their X-ray emission in terms of proton synchrotron radiation,
rather than as the inverse-Compton boosted cosmic-microwave background
photons \citep[e.g.,][]{2002MNRAS.332..215A,2004ApJ...613L..25H,2006ApJ...648..910U,2016ApJ...817..121B,2017ApJ...849...95B}.
The needed high energy protons could even be injected into the jet
via turbulent acceleration in the shear layer of the jet \citep{2000MNRAS.312..579O}.
Alternatively, the x-ray jets could be synchrotron radiation produced
by the secondary electrons arising from photo-hadronic interactions
like the Bethe-Heitler process or photopion production \citep[see][]{2017MNRAS.464.2213P,2017ApJ...835...20K,2016ApJ...817..121B}.
In all such scenarios of hadronic jets, one would expect neutrino
emission from charged pion or neutron decay resulting from interactions
between high energy protons and photons \citep{1989A&A...221..211M}.
The recent observation of a high energetic neutrino from the Blazar
TXS 0506+056, is best explained by a hadronic emission process \citep{2018arXiv180704300A}.
A similar signal has not yet been observed from FR II radio galaxies,
which may constrain the hadronic jet model \citep{2017A&A...603A.135N}.

Lepto-hadronic model for high-energy emission from FR I radio galaxies
are consistent with their observed high energy radiation \citep{article}.

Over the years, significant observational evidence has in fact emerged
in support of a dominant leptonic relativistic plasma component in
the extended radio lobes of FR II sources. The evidence has come mainly
from balancing the observationally estimated pressure of the external
X-ray emitting thermal plasma and the internal lobe pressure due to
the nonthermal plasma estimated by modeling of the combined synchrotron/inverse-Compton
spectrum of the lobe, as constrained by measurements at radio and
X-ray energies, respectively (e.g.,\citep{2017MNRAS.467.1586I,2002ApJ...581..948H,2004MNRAS.353..879C,2013MNRAS.436.1595K}).
This concurs with the conclusion reached indpendently for the lobes
of several FR II sources, including Cygnus A, based on modeling of
the observed shapes of the radio lobes (see \citep{2016MNRAS.457.1124K}).
The X-ray emission from hotspots of powerful FR II sources is often
consistently explained in terms of the SSC mechanism, when the magnetic
field is close to the equipartition value and under the assumption
of an energetically significant, if not a dominant, pair-plasma (vis
a vis the electron-proton plasma) \citep{2004ApJ...612..729H,2005ApJ...622..797K},
which is in accord with the findings for powerful radio lobes of FRII
sources (see above). Specifically, for the well studied case of Cygnus
A, the analysis of the X-ray emission from its hotspots has lent strong
support to such an interpretation \citep{2004ApJ...612..729H,2000ApJ...544L..27W}.
Clearly, pair-plasma dominated powerful hotspots would be entirely
in tune with the dominance of pair-plasma inferred for FR II lobes
(see above), since the latter are fed by the hotspots. Nonetheless,
given the various uncertainties involved, such as spatial inhomogeneities
in the radiating plasma and beaming of the radiation, independent
observational constraints on the presence of an energetic proton population
in the hotspots would be very desirable. A population of energetic
protons in hot spots could be the origin of the ultra-high energetic
cosmic rays observed, as the hot spots of FR II radio galaxies are
potential acceleration sites (see e.g. \citep{2018JCAP...02..036E}
and reference therein). They would also be in line with the recent
detection of neutrinos from the Blazar TXS 0506+056 \citep{2018arXiv180704300A}.

There is a long history of circular polarization measurements aiming
to determine the relativistic plasma composition near the jet's base
itself. More recently, VLBI imaging of both circular and linear polarization
have been carried out for a few blazars on sub-parsec scale, sometimes
at multiple radio frequencies \citep{2009ApJ...696..328H,2008MNRAS.384.1003G}.
For the blazar 3C 279, Homan et al. \citep{2009ApJ...696..328H} have
reported a particularly detailed observational study which also includes
simulations of their sensitive VLBI results in Stokes I, linear polarization,
and circular polarization at 6 frequencies in the range 8 to 25 GHz.
The significant detection of CP is thus interpreted by them primarily
in terms of Faraday conversion of the linear polarization within the
nuclear jet. Their main conclusion is that the jet is kinetically
dominated by electron-proton plasma, though a significant presence
of pair-plasma may still contribute to the radiation. Clearly, even
this detailed analysis is vulnerable to uncertainties in the jet's
physical parameters, like bulk speed and the minimum Lorenz factor
of the relativistic plasma ($\gamma_{\mathrm{min}}$). Interestingly,
these uncertainties are largely obviated in the case of Cygnus A hotspots
since not only is their motion non-relativistic (like all hotspots
on kiloparsec-scale) but estimates of $\gamma_{\mathrm{min}}\sim600$
are also available, based on recent spectral turnover measurements
made with the LOFAR telescope \citep{2016MNRAS.463.3143M}.

The energy of the leptons that are visible within the observationally
accessible radio wavebands is usually not sufficient to explain the
rough pressure balance between radio lobes and their surrounding thermal
plasma. Additional relativistic protons could fill in the deficit.
However, a large population of leptons with lower than observable
energies could as well bridge the gap, as would deviations from the
usually invoked equipartition assumption between particles and magnetic
fields. For these reasons, a more direct determination of the composition
of radio plasma via CP measurements as proposed here would be very
important.

\section{Estimation\label{sec:Estimation}}

In the following we estimate the expected fractional CP emission arising
from the hotspots of Cygnus A, in the case of a pure relativistic
electron-proton plasma. A volume element of a hotspot along a line
of sight (LOS) may harbor $n_{\mathrm{e}}$ relativistic electrons
with a power law spectrum
\begin{align}
\frac{dn(\gamma)}{d\gamma} & =\frac{n_{\mathrm{e}}\gamma^{-p}}{\gamma_{\mathrm{min}}^{1-p}-\gamma_{\mathrm{max}}^{1-p}}
\end{align}
between $\gamma_{\mathrm{min}}$ and $\gamma_{\mathrm{max}}$ and
with energy spectral index $p$. Their synchrotron emissivity is 
\begin{eqnarray}
j & = & \begin{pmatrix}j_{I}\\
j_{Q}\\
j_{U}\\
j_{V}
\end{pmatrix}=\underbrace{j_{0}n_{\mathrm{e}}B_{\perp}^{\frac{p+1}{2}}\,\nu^{\frac{1-p}{2}}}_{j_{I}}\,\begin{pmatrix}1\\
q\\
u\\
v
\end{pmatrix},
\end{eqnarray}
where
\begin{eqnarray}
j_{0} & = & \frac{\mathrm{e}^{2}}{c}\,\left(\frac{\mathrm{e}}{2\pi m_{\mathrm{e}}c}\right)^{\frac{p+1}{2}}\frac{3^{\frac{p}{2}}(p-1)\,\Gamma\left(\frac{3p-1}{12}\right)\,\Gamma\left(\frac{3p+19}{12}\right)}{2\,(p+1)\,(\gamma_{\mathrm{min}}^{1-p}-\gamma_{\mathrm{max}}^{1-p})}\\
q & = & -\frac{p+1}{p+\nicefrac{7}{3}}\cos\left(2\varphi\right)\\
u & = & +\frac{p+1}{p+\nicefrac{7}{3}}\sin\left(2\varphi\right)\\
v & = & -\frac{j_{1}\,B_{\|}}{\left(\nu\,B_{\perp}\right)^{\nicefrac{1}{2}}}\\
j_{1} & = & \frac{171}{250}\,\left(\frac{3\,\mathrm{e}\,p}{2\pi m_{\mathrm{e}}c}\right)^{\nicefrac{1}{2}}=0.06268\,\left(\frac{p\,\mbox{GHz}}{\mbox{Gauss}}\right)^{\nicefrac{1}{2}}
\end{eqnarray}
and the magnetic field is 
\[
\mathbf{B}=\begin{pmatrix}B_{\perp}\cos\left(\varphi\right)\\
B_{\perp}\sin\left(\varphi\right)\\
B_{\parallel}
\end{pmatrix}=\begin{pmatrix}\sin(\theta)\,\cos\left(\varphi\right)\\
\sin(\theta)\,\sin\left(\varphi\right)\\
\cos(\theta)
\end{pmatrix}\,B,
\]
with the $z$-axis being parallel to the LOS and $\varphi$ the angle
between the $x$-axis and the field component in the plane of the
sky, $\mathbf{B}_{\perp}$\citep{2016ApJ...822...34P}. The observational
frequency is denoted by $\nu$. The symbols $\mathrm{e},\,m_{\mathrm{e}}$
and $c$ denote the elementary charge, the electron mass, and the
speed of light, respectively. Relativistic positrons have exactly
the same emissivities, with the only difference of an opposite Stokes
$V$ sign. For this reason, there is no CP synchrotron emission in
case the electron and positron spectra are identical in slope and
normalization.

The total polarized emission of our electron-proton plasma is the
volume integrated emissivity, 
\begin{equation}
J=\int_{\mathcal{V}}dx\,j(x)\approx\sum_{i=1}^{N_{\mathrm{cell}}}\mathcal{V}_{i}\,j_{i},
\end{equation}
which we assume to be made of $N_{\mathrm{cell}}$ similar sized cells
with spatially constant emissivity $j_{i}$ within the cell volume
$\mathcal{V}_{i}\approx\mathcal{V}_{\mathrm{cell}}.$ 

For simplicity, we assume only the magnetic field orientation to vary
from cell to cell, whereas the field strength $B$ and the electron
spectrum $\frac{dn(\gamma)}{d\gamma}$ are taken to be approximately
the same within the entire emitting volume $\mathcal{V}=N_{\mathrm{cell}}\mathcal{V}_{\mathrm{cell}}$.
As no magnetic field direction is a priori preferred, we assume a
random distribution of field orientations from cell to cell and therefore
get for the average, dispersion, and root mean square (rms) of a quantity
$X$
\begin{eqnarray}
\overline{X} & = & \frac{1}{4\pi}\int_{0}^{2\pi}d\varphi\int_{0}^{\pi}d\theta\,\sin\theta\,X(\theta,\varphi),\\
\sigma_{X}^{2} & = & \frac{\overline{(X-\overline{X})^{2}}}{N_{\mathrm{cell}}},\mbox{ and }\\
X_{\mathrm{rms}} & = & \sqrt{\overline{X^{2}}}=\sqrt{\overline{X}^{2}+\sigma_{X}^{2}}.
\end{eqnarray}

The expected mean of the emission is $\overline{J}=(\overline{I},\,0,\,0,\,0)^{\mathrm{t}}$,
with $\overline{I}=\overline{j_{I}}\mathcal{V}$ and $\overline{j_{I}}\approx\,0.719j_{0}n_{\mathrm{e}}B^{\frac{3}{2}}\,\nu^{\frac{-1}{2}}$ for $p=2$.
For an individual emission region like a hot spot, the different Stokes
parameters show nonzero variance:

\begin{equation}
\begin{pmatrix}\sigma_{I}\\
\sigma_{Q}\\
\sigma_{U}\\
\sigma_{V}
\end{pmatrix}=\begin{pmatrix}0.190\\
0.106\\
0.106\\
0.0229\,B_{\mathrm{Gauss}}^{\nicefrac{1}{2}}\,\nu_{\mathrm{GHz}}^{-\nicefrac{1}{2}}
\end{pmatrix}\,\frac{j_{0}n_{\mathrm{e}}B^{\frac{3}{2}}\,\nu^{\frac{-1}{2}}\mathcal{V}}{\sqrt{N_{\mathrm{cell}}}}
\end{equation}

One should note that the assumption of randomness in the directional
distribution is related to the cell sizes and the typical coherence
length of the magnetic field, which may be rather large in Cygnus
A due to the presence of shocks. In the most extreme limit of a completely
coherent magnetic field one may set $N_{\mathrm{cell}}=1$, in order
to get a intuition for this effect. 

In any case, the cell model is just a simplification, which may be
dropped for a more elaborate calculation based on a physical model.
It should, however, provide the right order of magnitude of the expected
effect in a statistical sense.

If we investigate the vector $J'=(I,\,P,V)^{\mathrm{t}}$ with $P=\sqrt{Q^{2}+U^{2}}$
being the total linear polarization, we find $\overline{J'}=(1,\frac{p+1}{p+\nicefrac{7}{3}}\,\frac{1}{N_{\mathrm{cell}}},\,0)^{\mathrm{t}}\overline{I}$,
since each cell has a fractional polarization of $\frac{p+1}{p+\nicefrac{7}{3}}\approx0.69$.
Thus, we expect 
\begin{equation}
\frac{\sigma_{V}}{\sigma_{P}}=\frac{\sigma_{V}}{\sqrt{\sigma_{Q}^{2}+\sigma_{U}^{2}}}=0.153\,B_{\mathrm{Gauss}}^{\nicefrac{1}{2}}\,\nu_{\mathrm{GHz}}^{-\nicefrac{1}{2}}.
\end{equation}
Given that the fractional linear polarization of the Cygnus A hot
spots is about $f_{P}=\sigma_{P}/\overline{I}\approx0.7$ \citep{1999AJ....118.2581C}
and and the field strength is $B\approx1.5\,10^{-4}\,\mbox{Gauss}$
\citep{1538-4357-544-1-L27}, we expect a fractional circular polarization
of about $f_{V}=\sigma_{V}/\overline{I}\approx1.31\cdot10^{-3}\,\nu_{\mathrm{GHz}}^{-\nicefrac{1}{2}}$.

In order to be able to correctly assess the expeced CP emission, we
must also consider the Faraday conversion effect of the foreground,
which translates linear into circular polarized emission.

From the generic radiative transfer equations we know that

\[
j_{V,\mbox{conv}}=\phi_{\text{c}}\lambda^{3}j_{Q},
\]
where $\text{\ensuremath{\phi}}_{\text{c}}$ is called the conversion
measure in analogy to the Faraday rotation measure $\phi_{\text{r}}$.
In the thermal regime it can be sufficiently fitted via \citep{2011MNRAS.416.2574H}

\[
\phi_{\text{c}}\approx\frac{\text{e}^{4}}{4\pi^{2}m_{\text{e}}^{3}c^{6}}\,\int_{\mathrm{LOS}}dr\ n_{\mathrm{th}}\ B_{\perp}^{2},
\]
under the assumption that the $j_{Q}$ emission takes place only in
the hotspots. The thermal electron density $n_{\text{th}}$ poses
a problem, as we do not have precise information on the electron column
density of the Cygnus cluster in the vicinity of the hotspots, where
most of the conversion is likely to take place. We do, however, have
precise measurements on the Faraday rotation $\phi_{\text{r}}$, which
contains the same $n_{\text{th}}$-dependence. We can write 

\[
\phi_{\text{c}}\approx\frac{\text{e}B_{\mathrm{ICM}}}{2\pi m_{\text{e}}c^{2}}\int_{\mathrm{LOS}}dr\frac{\text{e}^{3}}{2\pi m_{\text{e}}^{2}c^{4}}n_{\mathrm{th}}B_{\text{ICM}}\approx\frac{\text{\ensuremath{\text{e}}}cB_{\mathrm{ICM}}}{2\pi m_{\text{e}}c^{2}}\phi_{\text{r}}
\]
under the assumption of a similar strength and correlation structure
of the intracluster magnetic field in both the parallel and perpendicular
component of the line of sight. We choose $B_{\mathrm{ICM}}=8\cdot10^{-6}\mathrm{G}$
as a typical value of the intracluster magnetic field \citep{1988ApJ...334L..73C}.
Cygnus A has very high values of rotation measures with around $\vert\phi_{\text{r}}\vert\approx\frac{\rho_{\phi}}{\lambda^{2}}\approx1000\,\frac{\mathrm{rad}}{\text{m}^{2}}$
\citep{1987ApJ...316..611D}, which we will choose for our estimate.
Therefore we arrive at $\phi_{c}\lambda^{3}\approx2.013\cdot10^{-6}\nu_{\mathrm{GHz}}^{-3}$. 

The mean $\overline{V_{\text{c}}}$ is again zero. For the variance
we find 
\[
\sigma_{V_{\text{c}}}=\phi_{\text{c}}\lambda^{3}\sigma_{Q}\approx2.133\cdot10^{-7}\frac{\overline{I}}{\nu_{\mathrm{GHz}}^{3}\sqrt{N_{\mathrm{cell}}}}
\]
 and for the ratio 
\[
\frac{\sigma_{V_{\text{c}}}}{\sigma_{V}}\approx9.32\cdot10^{-6}\frac{1}{B_{\mathrm{Gauss}}^{\nicefrac{1}{2}}\nu_{\mathrm{GHz}}^{2.5}}.
\]
Although small in the $\text{\ensuremath{\mathrm{GHz}}}$ regime,
the above ratio reaches unity rather fast for smaller frequencies.
We expect $\sigma_{V_{\text{c}}}/\sigma_{V}\approx1$ for $\nu\approx57$
MHz and the aforementioned hotspot magnetic field strength of $B\approx1.5\,10^{-4}\,\mbox{Gauss}.$
Therefore, observations well above 56 MHz should not be significantly affected by Faraday conversion.

\section{Conclusions\label{sec:Conclusions}}

An electron-proton jet is expected to give rise to a fractional CP
at a level of $10^{-3}(\nu/\mbox{GHz})^{-\nicefrac{1}{2}}$. Clearly,
this is very challenging to measure, given the faintness of the signal
and the systematic cross talk of polarized radio receivers. Nevertheless,
systematic effects change from receiver to receiver, and even as a
function of time as the relative orientation of sky and the telescope
changes with the Earth's rotation. This engenders some hope that genuine
CP detection might become technically feasible.

Due to the effect of Faraday conversion of LP into CP, we expect the
optimal frequency window for the detection of CP to be around $70\,\mathrm{MHz}.$
This renders low frequency radio telescopes like LOFAR most suitable.
However, an excellent polarimetry is important as well, so that the
the JVLA and in future the Square Kilometer Array as well as its precursors,
the MeerKAT and ASKAP telescopes, are promising. The necessary polarimetric
accuracies have already been achieved in observations \citep{doi:10.1111/j.1365-8711.2000.03854.x}.

A detection of the CP signal and thereby a confirmation of the electron-proton
jet scenario might therefore be in reach soon. This would be very
exciting, as it would contradict the prevailing majoritarian view
in the radio galaxy community on jet composition in FR II radio galaxies.
At the same time, it would make it more plausible that such galaxies
are the acceleration sites of ultra high energy cosmic rays and it
would also be in line with the recent detection of neutrinos from
the Blazar TXS 0506+056 \citep{2018arXiv180704300A}. The exclusion
of electron-positron jets would, however, not be absolute. A possible
scenario in which such jets can produce CP is the case of differing
electron and positron energy spectra. These could be caused by their
acceleration in a charge asymmetric environment, for example due to
the entrainment of thermal electron-proton plasma into the particle
acceleration sites, so that plasma physical effects might engender
different acceleration efficiencies of electrons and positrons. 

A non-detection with a sensitivity well below the CP flux level estimated
here for the hotspots of Cygnus A would also be very interesting,
as it would argue strongly in favour of an electron-positron pair
plasma, where both electrons and positrons have similar energy spectra,
indicating a charge symmetric genesis and acceleration history for
both species. The exclusion of the electron-proton jet scenario would
be relatively firm in this case, as the only remaining way to avoid
predicting detectable CP emission would be that the magnetic field
in the hotspots points in a highly matched fashion towards and away
from the observer, such that the CP emission of those volumes cancel.
Given the high linear polarisation of the Cygnus A hotspots, and therefore
an imbalance of the field orientations projected on the plane of the
sky, a symmetry along the line of sights seems unlikely. Spatially
resolved CP observations, or the study of a larger sample of hotspots
of FR II galaxies could negate such an explanation.

To conclude, we have shown that sensitive CP observations of radio
galaxy hotspots are a promising way to determine the composition of
synchrotron plasma in radio galaxies in a way which is independent
of many of the assumptions made in other lines of investigation probing
this question. 
\begin{acknowledgments}
We would like to thank the organizers of the workshop \textit{Plasma
universe and its structure formation} at the Inter-University Centre
for Astronomy and Astrophysics, Pune, India in 2017 for the stimulating
atmosphere that led to this research. Furthermore, we thank an anonymous
referee for constructive comments and Richard Anantua and Razieh Emami for pointing out an error in the original manuscript that is corrected here.
\end{acknowledgments}

\appendix
\bibliographystyle{apsrev}
\bibliography{CPol}

\end{document}